\documentclass[useAMS,usenatbib]{mn2e}

\usepackage{amsmath}
\usepackage{graphics}
\usepackage{float} 
\usepackage{array}
\usepackage{color} 
\definecolor{Darkgreen}{rgb}{0,0.4,0}
\definecolor{Darkred}{rgb}{0.8,0,0} 
\definecolor{Darkblue}{rgb}{0.0, 0.0, 0.55}
\usepackage[colorlinks=true,urlcolor=blue,linkcolor=Darkblue,filecolor=black,citecolor=Darkblue]{hyperref}
\usepackage{graphicx}
\usepackage{natbib}
\usepackage[final]{pdfpages} 
\usepackage{hyperref}
\usepackage{pdflscape}
\usepackage{lscape}
\usepackage{rotating}
\usepackage[section]{placeins}
\usepackage{xcolor,colortbl}
\usepackage{placeins}
\usepackage{dblfloatfix}

\newcommand{\deriv}{\mathrm{d}}

\definecolor{Gray}{gray}{0.85}

\title{Testing the accuracy of clustering redshifts with simulations}

\author[V.~Scottez et al.]{V.~Scottez$^{1}$\thanks{E-mail:
scottez@iap.fr}, A.~Benoit-L\'evy$^{1}$, J.~Coupon$^{2}$, O.~Ilbert$^{3}$, Y.~Mellier$^{1,4}$\\
$^{1}$Institut d'Astrophysique de Paris, UMR7095 CNRS, Universit\'e Pierre \& Marie Curie, 98 bis boulevard Arago, 75014 Paris, France\\
$^{2}$Department of Astronomy, University of Geneva, ch. d'\'Ecogia 16, 1290 Versoix, Switzerland\\
$^{3}$Aix Marseille Universit\'e, CNRS, LAM (Laboratoire d'Astrophysique de Marseille) UMR 7326, 13388, Marseille, France\\
$^{4}$CEA/Irfu/SAp Saclay, Laboratoire AIM, F-91191 Gif-sur-Yvette, France}

\begin{document}

\def\aj{AJ}%
\def\actaa{Acta Astron.}%
\def\araa{ARA\&A}%
\def\apj{ApJ}%
\def\apjl{ApJ}%
\def\apjs{ApJS}%
\def\ao{Appl.~Opt.}%
\def\apss{Ap\&SS}%
\def\aap{A\&A}%
\def\aapr{A\&A~Rev.}%
\def\aaps{A\&AS}%
\def\azh{AZh}%
\def\baas{BAAS}%
\def\bac{Bull. astr. Inst. Czechosl.}%
\def\caa{Chinese Astron. Astrophys.}%
\def\cjaa{Chinese J. Astron. Astrophys.}%
\def\icarus{Icarus}%
\def\jcap{J. Cosmology Astropart. Phys.}%
\def\jrasc{JRASC}%
\def\mnras{MNRAS}%
\def\memras{MmRAS}%
\def\na{New A}%
\def\nar{New A Rev.}%
\def\pasa{PASA}%
\def\pra{Phys.~Rev.~A}%
\def\prb{Phys.~Rev.~B}%
\def\prc{Phys.~Rev.~C}%
\def\prd{Phys.~Rev.~D}%
\def\pre{Phys.~Rev.~E}%
\def\prl{Phys.~Rev.~Lett.}%
\def\pasp{PASP}%
\def\pasj{PASJ}%
\def\qjras{QJRAS}%
\def\rmxaa{Rev. Mexicana Astron. Astrofis.}%
\def\skytel{S\&T}%
\def\solphys{Sol.~Phys.}%
\def\sovast{Soviet~Ast.}%
\def\ssr{Space~Sci.~Rev.}%
\def\zap{ZAp}%
\def\nat{Nature}%
\def\iaucirc{IAU~Circ.}%
\def\aplett{Astrophys.~Lett.}%
\def\apspr{Astrophys.~Space~Phys.~Res.}%
\def\bain{Bull.~Astron.~Inst.~Netherlands}%
\def\fcp{Fund.~Cosmic~Phys.}%
\def\gca{Geochim.~Cosmochim.~Acta}%
\def\grl{Geophys.~Res.~Lett.}%
\def\jcp{J.~Chem.~Phys.}%
\def\jgr{J.~Geophys.~Res.}%
\def\jqsrt{J.~Quant.~Spec.~Radiat.~Transf.}%
\def\memsai{Mem.~Soc.~Astron.~Italiana}%
\def\nphysa{Nucl.~Phys.~A}%
\def\physrep{Phys.~Rep.}%
\def\physscr{Phys.~Scr}%
\def\planss{Planet.~Space~Sci.}%
\def\procspie{Proc.~SPIE}%
\let\astap=\aap
\let\apjlett=\apjl
\let\apjsupp=\apjs
\let\applopt=\ao

\uchyph=0

\newcommand{\vs}[1]{\textcolor{red}{ #1}}

\date{Accepted 2017 ... . Received 2017 ... ; in original form 2017 ...}

{\pagerange{\pageref{firstpage}--\pageref{lastpage}} \pubyear{2017}

\maketitle

\label{firstpage}

\begin{abstract}
We explore the accuracy of the clustering-based redshift inference within the MICE2 simulation. This method uses the spatial clustering of galaxies between a spectroscopic reference sample and an unknown sample. The goal of this study is to give a preview of the redshift accuracy one can reach with this method. To do so, we first highlight the requirements of this technique in term of number of objects in both the reference and unknown samples. We also confirm that this method does not require a representative spectroscopic sample for calibration. 

We estimate that a density of spectroscopic objects of $10^{-5}$ arcmin$^{-2}$ per redshift bin of width $\delta z = 0.01$ over $9000 \ \text{deg}^{2}$ allows to reach 0.1 \% accuracy in the mean redshift for a galaxy density compatible with next generation of cosmological surveys. This number is compatible with the density of the Quasi Stellar Objects in BOSS. Second we demonstrate our ability to measure individual redshifts for galaxies independently from the photometric redshifts procedure. The resulting individual clustering redshifts have a bias=$-0.001$, an outlier fraction of $\eta=3.57\%$ and a scatter of $\sigma=0.027$ to $i<25$. The advantage of this procedure is threefold: i)  it allows the use of clustering redshifts for any field in astronomy, ii) it allows the possibility to combine photometric and clustering based redshifts to get an improved redshift estimation,  iii) it allows the use of cluster-$z$ to define tomographic bins for weak lensing. Finally we explore this last option and build 5 clustering redshift selected tomographic bins from redshift 0.2 to 1. We found a bias on the mean redshift estimate of $0.002$ per bin.

\end{abstract}


\section{Introduction}
\label{sec:intro}
The next generation of cosmological surveys such as Euclid \citep{euclid_mission,amendola_2012} , LSST \citep{lsst_mission2} and WFIRST \citep{wfirst_mission}, whose aim is to understand the largescale accelerated expansion of the universe will perform surveys of the sky. One of the major issues in astronomy is to be able to deproject the 2D information into a 3D one.

At extragalactic scales, the study of the universe relies on our knowledge of the redshift as a tracer of distances. The ideal way of measuring resdhifts is to do spectroscopie and measure their spectroscopic redshift (spectro-$z$). While this method is the most accurate, it is also very time-consumming. Another largely used approach is to estimate the redshift of an object based on its flux information through some filters. This is the so called photometric redshift (photo-$z$) procedure which require large and representative set of spectroscopic redshift measurements for the calibration of empirical methods \citep{Connolly_1995} and the building of representative template libraries for template-fitting techniques \citep{coleman_1980}. This procedure uses the correlation between the colours of the population of unknown redshift and the flux from a reference sample. The photo-z procedure is fast with respect to the spectroscopic one and allows to measure redshifts for millions of objects. One difficulty is that current and future spectroscopic surveys are incomplete in magnitude, redshift and galaxy properties \citep{cooper_2006,newman_2013_spectro_needs}, which is challenging to calibrate photometric redshifts \citep{masters_2017}. Also, projections for cosmic shear measurements estimate that the true mean redshift of objects in each photo-z bin, considering Euclid statistic, must be known to better than $\Delta \langle z \rangle = 0.002(1+z)$ \citep{knox_2006,zhan_2006,zhan_and_knox_2006}.\newline
\indent The method we used in this study relies on the clustering properties of galaxies. This approach is looking at the spatial correlation between galaxies at unknwon redshift and reference sources with known spectroscopic redshift. Fundamentaly this approach only require the knowledge of unknown galaxies on-sky position.\newline
\indent The idea of measuring redshift distributions using the apparent clustering of objects was first developed by \cite{seldner_peebles_1979};
\cite{phillipps_shanks_1987} and \cite{landy_szalay_1996}. This was
practically forgotten mainly due to the rise of photometric
redshifts. To face the challenges of future and ongoing dark energy
imaging experiment \cite{newman_2008}, \cite{mattew_newman_2010} and
\cite{matthews_newman_2012} used the large scale clustering of galaxies and iteratively approximated the galaxy to dark matter bias and validated this method on simulations
while \cite{mcquinn_white_2013} proposed an optimal estimator for such
a measurement.\newline
\indent We use the method presented in \cite{menard_2013}\& \cite{schmidt} compared to spectroscopic redshift by \cite{rahman_2015}. This method focused on lower scales, where the clustering signal is higher, and assume that there is no or little galaxy to dark matter bias evolution when the redshift distribution is narrow. This approach was also applied to continuous fields by inferring the redshift distribution of the cosmic infrared background \cite{schmidt_2015} while \cite{rahman_2015_IR} and \cite{rahman_2015_sdssphotz} explored this method for near infrared data using 2MASS Extended and Point Source Catalogs as well as the SDSS Photometric Galaxies. Recently a similar approach was used to validate the redshift distribution used for weak lensing by \cite{kids_450}.\newline
\indent The goal of this study is to give a preview of the redshift accuracy one can reach with this method. For this, we explore the accuracy and the requirements of this technique. We also present  individual clustering-based redshift measurement for a single galaxy using the MICE2 simulation. Finally we investigate the use of this method to define tomographic bins for weak lensing by building 5 clustering redshift selected bins from redshift 0.2 to 1.\newline
\newline
\indent This paper is organised as follows. In Section \ref{sec:cluster_z_formalism} we review the clustering-based redshift formalism, while the MICE2 simulation and the data used to build the reference and the unknown samples are introduced in Section \ref{sec:Euclid_sim}. In Section \ref{sec:systematic_tests} we explore the evolution of the clustering redshift distribution accuracy when varying (1) the magnitude of the unknown sample, (2) the number of unknown and reference sources as well as (3) the width of the unknown distribution. Based on these results we estimate the requirements in terms of spectroscopic data needed per redshift bin for the next generation of cosmological surveys. Then we demonstrate our ability to measure clustering redshifts for individual galaxies and quantify their accuracy, in Sections \ref{sec:individual_redshift}. Based on these measurements, we investigate the possibility to directly use the clustering redshift to select  tomographic bins for weak lensing. Finally, we present 5 clustering redshift selected tomographic bins for which we give an estimate of the bias. Conclusions are presented in Section \ref{sec:summary}.

\section{Clustering redshift formalism}
\label{sec:cluster_z_formalism}
In this section we briefly review the clustering-based redshift formalism initially developed in \cite{menard_2013}, \cite{rahman_2015} and fully detailed in \cite{scottez_phd_2015}.

The integrated cross-correlation can be written as:
\begin{equation}
\bar{\omega}_{\text{ur}}(z) = \int \deriv z' \frac{\deriv N_{\text{u}}}{\deriv z'} \frac{\deriv N_{\text{r}}}{\deriv z'} \beta_{\text{u}}(z') \beta_{\text{r}}(z') \ ,
\end{equation}
\label{eqomega_bar_definition}
\noindent where $\deriv N_{\text{i}} / \deriv z$ are the redshift distributions, $\beta_{\text{i}}(z)$ the clustering amplitude with $i \in \left\{ \text{u,r} \right\}$ where lower script $\text{u,r}$ refers to the unknown and reference samples respectively. Considering a narrow sampling of the reference population one can write:
\begin{equation}
\frac{\deriv N_{\text{u}}}{\deriv z} = \bar{\omega}_{\text{ur}}(z) \times \frac{1}{\beta_{\text{u}}(z)}  \times \frac{1}{\beta_{\text{r}}(z)} \times \frac{1}{N_{\text{r}}} \ .
\end{equation}
\label{eqdn_dz_full}
This expression requires the knowledge of $\beta_{\text{u}}(z)$ which is a priori not known. There are two ways to bypass this issue. First, one can estimate this term by measuring the auto-correlation functions of both the unknown and reference populations as described in \cite{newman_2008} and \cite{kids_450}. Or, one can build unknown subsamples localised in redshift and then consider no \-- or linear \-- evolution for $\beta_{\text{u}}(z)$ (see \cite{schmidt}, \cite{rahman_2015} and \cite{scottez_vipers_2016}) which leads to:
\begin{equation}
\frac{\deriv N_{\text{u}}}{\deriv z} \propto \bar{\omega}_{\text{ur}}(z)  \times \frac{1}{\beta_{\text{r}}(z)} \ .
\end{equation}
\label{eqdn_dz_to_be_normed}
Where $\deriv N_{\text{u}} / \deriv z$ needs to be normalised to the number of objects in the unknown subsample.

\section{The MICE simulation}
\label{sec:Euclid_sim}

\noindent The MICE simulation catalog \citep{mice2_1,mice2_2,mice2_3,mice2_4} was generated using a hybrid Halo Occupation Distribution (HOD) and halo Abundance Matching (HAM) prescriptions to populate Friends of Friends (FOF) dark matter halos from the MICE-Grand Challenge (MICE-GC) simulation.

The catalog used as input the light-cone of the MICE-GC N-body run simulation. The input cosmological parameters are $\Omega_{\text{m}}=0.25$, $\sigma_{\text{8}}=0.8$, $n_{\text{s}}=0.95$, $\Omega_{\text{b}}=0.044$, $\Omega_{\Lambda}=0.75$, $h=0.7$. Further details on the simulation can be found at the MICE webpage (\url{www.ice.cat/mice}).

The catalog was built to follow local observational constraints:
\begin{itemize}
\item The local luminosity function (\cite{blanton_2003}   and \cite{blanton_2005} for the faintest galaxies).
\item The galaxy clustering as a function of luminosity and colour \citep{zehavi_2011}.
\item The color-magnitude diagram (NYU dr7 catalog).
\end{itemize}
\noindent It also includes:
\begin{itemize}
\item galaxy evolution in order to better match the luminosity function and the colour distributions at high redshift.
\item magnified positions due to gravitational lensing effects computed using projected mass density maps (in HEALPIX format) of the MICE-GC light-cone simulation.
\end{itemize}

\noindent We apply the following evolutionary correction for all magnitudes to get the right number counts at higher redshifts:

\begin{equation}
\text{mag}_{\text{ev}} = \text{mag}_{\text{cat}} - 0.8  \times (\arctan(1.5 \times z_{\text{cgal}} ) - 0.1489) \ ,
\label{eqmag_correction}
\end{equation}
\noindent where $\text{mag}_{\text{ev}}$ is the magnitude corrected for the evolutionary correction, $\text{mag}_{\text{cat}}$ is the original one (that does NOT include any correction) and $z_{\text{cgal}}$ is the true redshift of the galaxy. In this paper, we explore and quantify the accuracy of clustering redshift using the state of the art simulation currently available. Nevertheless, because of the incompleteness at $irz\sim 24.5$ and the limited redshift range to $z<1.4$, it is not possible to create, for example, a Euclid like scenario.

\subsection{Data selection}

\noindent In this exploratory work, since our goal is not to explore the maximum limit accuracy of this method nor to reach any given requirement but to give a preview of the achievable accuracy for individual redshift and mean redshift of a distribution, we choose for simplicity to work on a subsample of $100$ deg$^{2}$ in the region $10^{\text{o}} < \text{RA} < 20^{\text{o}}$, $10^{\text{o}} < \text{Dec} < 20^{\text{o}}$. The application to a larger area for a given survey and the scientific application correspond to a full study that we will investigate in a next paper. The parent sample results in $8.5 \text{M}$ galaxies for which we corrected the magnitudes according to Equation \ref{eqmag_correction} and for which we selected the following fields: RA,DEC, ZSPEC and ugrizYJH-bands. This parent sample will be used to build both the reference and the unknown samples.

\begin{figure}
\begin{center}
\includegraphics[scale=0.35]{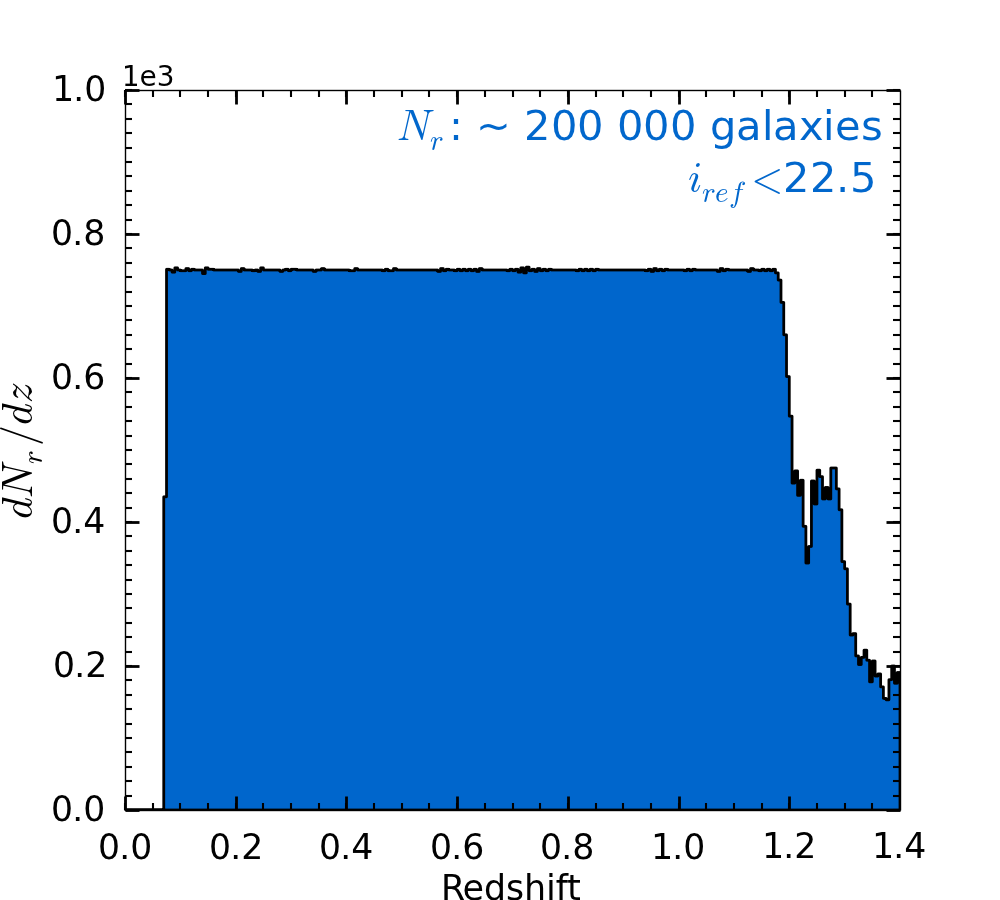}
\end{center}
\caption{Redshift distribution of the reference sample. The decreasing number of sources from $z=1.2$ directly reflect the lack of sources in the MICE2 simulation at these redshifts for $i<22.5$.}
\label{fig_ref_nz}
\end{figure}

\newpage
\subsubsection{Reference sample}

We choose to build an optimistic reference sample made of $\sim 200 \ 000$ sources corresponding to $\sim 0.55$ gal.arcmin$^{-2}$. These sources are randomly selected from the parent sample to have $i_{\text{r}}<22.5$ and a flat distribution in redshift as show in Figure~\ref{fig_ref_nz}. This choice allows us to reduce statistical fluctuation in the detection due to the variation in the number of sources over the range of references slices. One can note that the number of reference sources decreases at high redshift. This is directly due to the decreasing number of sources available with $i<22.5$ at these redshifts from the MICE2 simulation. Therefore, we do expect a higher noise level between redshift 1.2 and 1.4.

\begin{figure}
\begin{center}
\includegraphics[scale=0.35]{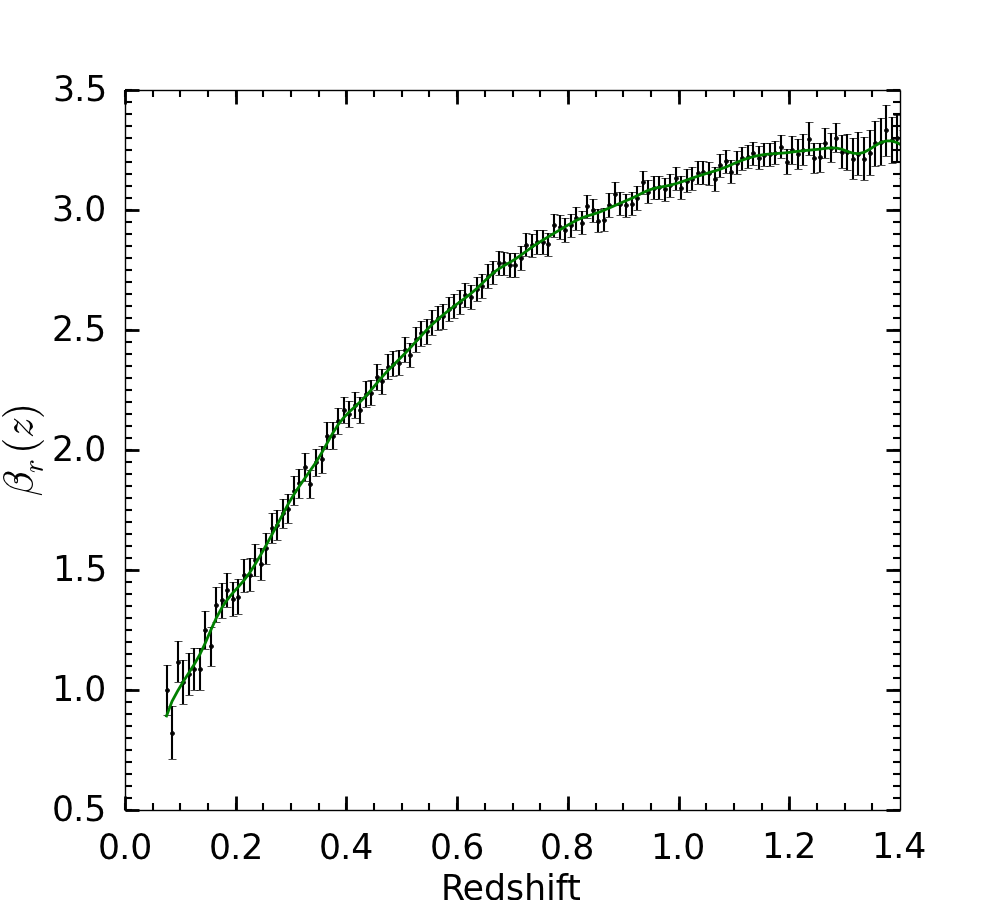}
\end{center}
\caption{Clustering amplitude of the reference population normalised to 1 at $z_0 = 0.07$. The solid green line shows a smoothed version of the measurement.}
\label{fig_beta_r}
\end{figure}

The evolution with redshift of the clustering amplitude of the reference sample $\beta_{\text{r}}$ is shown in Figure~\ref{fig_beta_r}. We choose to normalize $\beta_{\text{r}}$ at $z_{\text{0}}=0.07$.

\begin{figure*}
\begin{center}
\includegraphics[scale=0.55]{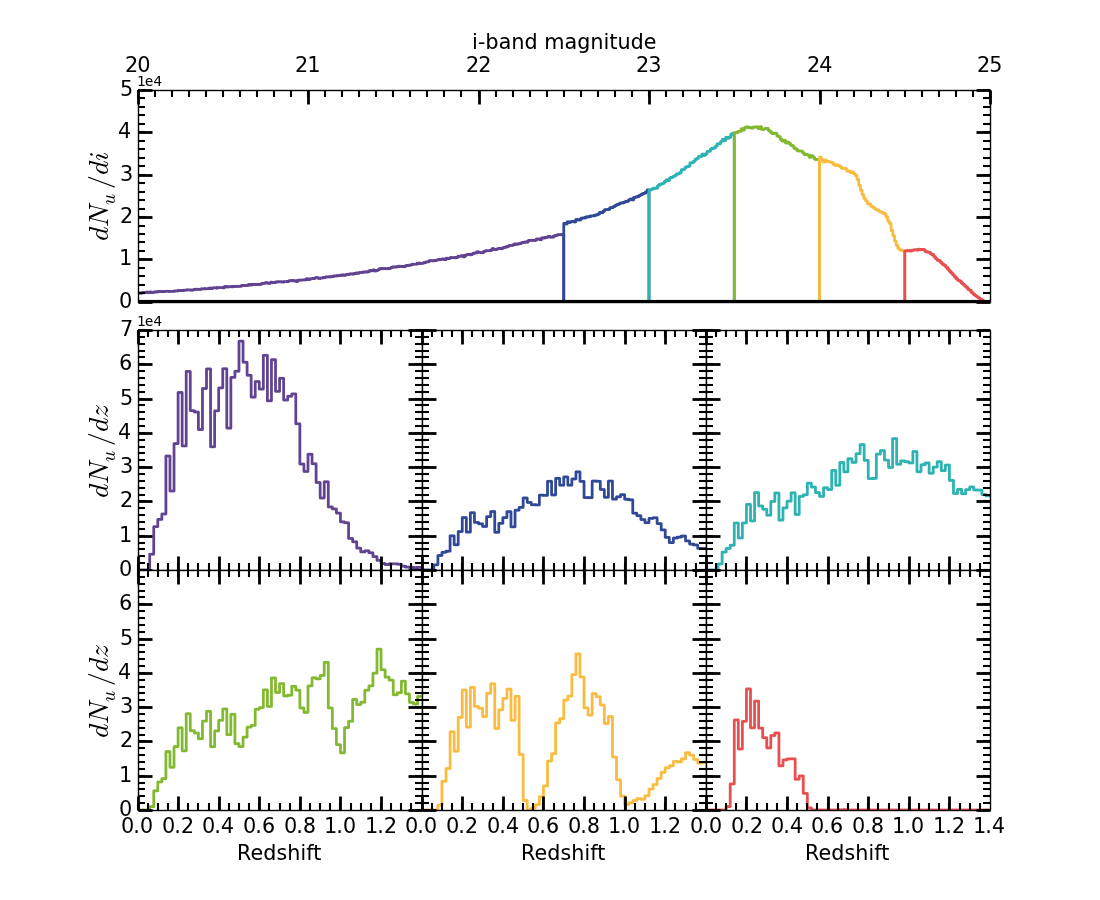}
\end{center}
\caption{\underline{TOP PANEL}:i-band magnitude distribution of the unknown sample. This sample starts to be incomplete at $i_u =23.5$. The break at $i_u =22.5$ is due to the objects removed from the parent sample to build the reference population.\newline
\underline{BOTTOM PANEL}: redshift distribution of each selected bin in i-band magnitude. The two fainter bins are incomplete in redshift due to the incompletness of the simulation at these magnitudes. From these 6 bins, we randomly select galaxies matching $\Delta z_{\text{u}}= 0.02$ and $N_{\text{u}}= 2 \ 000$ to build the tomographic subsamples.}
\label{fig_vary_with_mag}
\end{figure*}

\subsubsection{Unknown sample}
\label{sec_unk_sample}

\noindent The unknown sample is made from the previous parent sample from excluding the sources selected to be in the reference population and contains $8.3 \text{M}$ galaxies corresponding to a density of $23$ gal.arcmin$^{-2}$. 

The top panel of Figure \ref{fig_vary_with_mag} shows the i-band magnitude distribution of the unknown sample.  This sample is complete to $i_u =23.5$. The following analysis is made using subsamples of this one. The break at $i_u =22.5$ is due to the objects removed from the parent sample to build the reference population. This figure also shows the redshift distributions for different bins in magnitude. The first bin in purple has $i_{\text{u}}<22.5$ and has the same magnitude limit than the reference sample. The next two bins (in blue and cyan) are fainter than the reference sample, but are still complete while the fourth bin (green) starts to be incomplete. Finally the last two magnitude bins (yellow and red) are more and more incomplete. This incompletness of the simulation for these samples induce unrealistic variations in redshift. This choice of binning will allow to study the evolution of clustering redshift accuracy with fainter magnitude and when the unknown sample is incomplete in Section \ref{sec:vary_with_mag}.

\subsection{Cleaning the reference sample}

\begin{figure*}
\begin{center}
\includegraphics[scale=0.4]{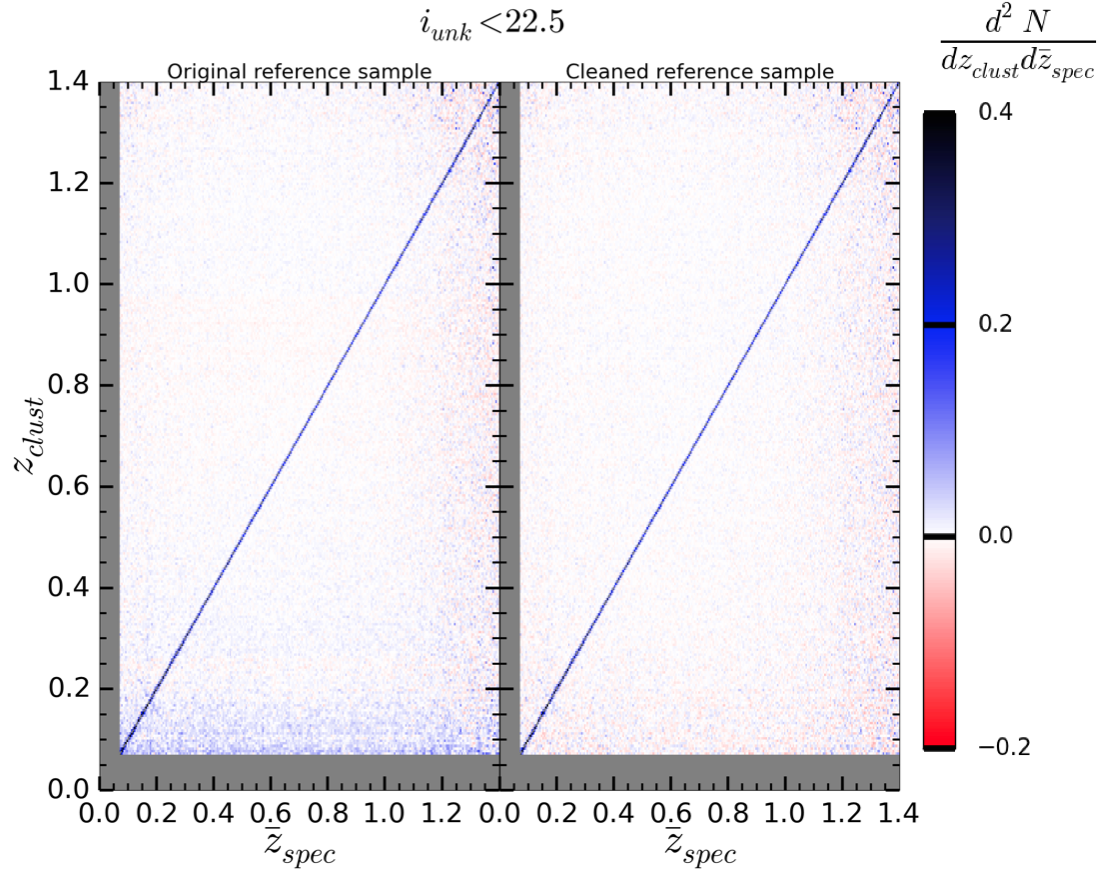}
\end{center}
\caption{\underline{LEFT PANEL}: clustering redshifts obtained with the full reference sample. One can note some horizontal stripes caused by the inhomogeneity of the reference sample. \newline
\underline{RIGHT PANEL}:  clustering redshifts obtained with the cleaned reference sample, where the stripes disappear.}
\label{fig_clean_ref_samp}
\end{figure*}

As highlighted by \cite{rahman_2015}, spurious correlation at low redshift ($z < 0.2$) can appear due to the inhomogeneous sampling arising from the clustered distribution of the objects in the reference sample. To address this issue, they removed all galaxies in regions where the measured densities are $2 \sigma$ from the mean. This procedure homogenises the spatial correlation of the reference sample, but also decreases the number of sources. To prevent spurious correlation at low redshift and optimise the statistical power of clustering-based redshift inference we weight each reference galaxy based on its local density.

To illustrate this, we select unknown tomographic subsamples according to the following criteria:
\begin{itemize}
\item true spectroscopic width of the unknown bin, $\Delta z_{\text{u}}= 0.005$.
\item number of reference sources, $N_{\text{r}}=200 \ 000$.
\item number of unknown objects in one bin, $N_{\text{u}}= \text{all available objects}$. Here, this number is not a constant. It can vary from bin to bin as shown in the middle left panel of Figure \ref{fig_vary_with_mag}.
\item magnitude of the unknown sample, $i_{\text{u}}<22.5$.
\end{itemize}
\noindent Then, for each unknown subsample we measure the corresponding clustering redshift distribution with and without cleaning the reference sample by individually weighting each reference galaxy by its local density. Both plots are shown in Figure \ref{fig_clean_ref_samp}.

The left panel shows the clustering redshift distributions obtained with the original reference sample. One can note the existence of horizontal stripes caused by the inhomogeneity of the reference sample. These stripes disappear in the right panel where we homogenize the reference population by individually weighting each reference galaxy by its local density. These weights are directly measured on the reference population and do not depend of the unknown sample. One can still note that the stochastic fluctuations beyond $z_{clust}=1.2$ are higher than bellow. This is due to the decreasing number of reference sources at these redshifts (see Figure \ref{fig_ref_nz}).

There is also a vertical stripe in both panels from $\bar{z}_{spec}$ 1.2 to 1.4. There are different effects at play in this region. First, this increased noise is mainly due to the low number of unknown objects per bin at these redshifts. Indeed, in the central part of the figure there is $\sim 10k$ unknown galaxies while in the vertical stripes between redshift 1.2 to 1.4 this number drops to $\sim 100$. Moreover, this range also corresponds to the redshift range where there are few reference galaxies, leading to a lower signal-to-noise ratio (SNR). To understand how each of these effects can affect the clustering based redshift estimation we present some tests in the next section.

\section{Tests on input parameters}
\label{sec:systematic_tests}

This section aims to highlight the behaviour of the clustering-based redshift inference with respect to the magnitude of the unknown sample, the number of objects in the unknown/reference samples and the width of the unknown distribution.

To do so, we consider an idealistic case in which we select galaxies of the unknown sample using theur true redshift to reduce the influence of the galaxies to dark matter bias. In each of the following subsections we fix a set of parameters and we vary one. The parameters are: the true width of the unknown tomographic bin $\Delta z_{\text{u}}$, the number of reference sources $N_{\text{r}}$, the number of unknown objects in the bin $N_{\text{u}}$, the magnitude of the unknown sample $i_{\text{u}}$. Each time, even when we vary $N_{\text{r}}$, we keep a flat redshift distribution for the reference sample.

\subsection{Evolution with the magnitude of the unknown sample: $i_{\text{u}}$ }
\label{sec:vary_with_mag}

\begin{figure*}
\begin{center}
\includegraphics[scale=0.4]{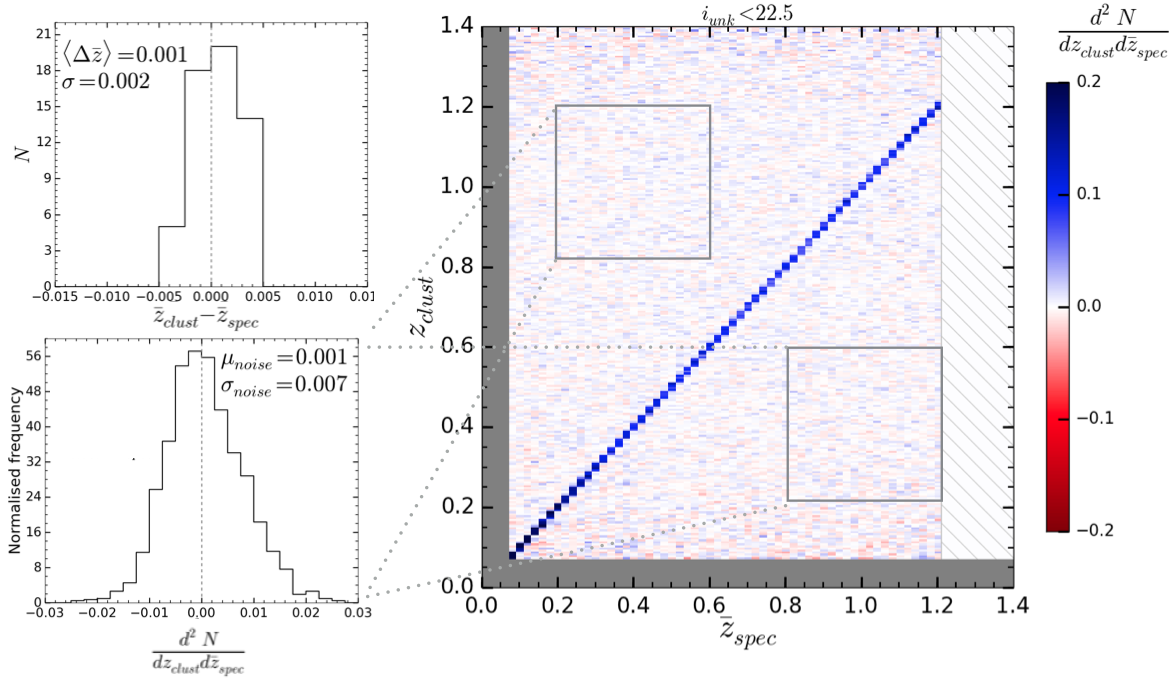}
\end{center}
\caption{Cluster/Spectro-z map for $i_{\text{u}}<22.5$. Regions where there is no galaxies available from the unknown sample are dashed. }
\label{fig_vary_mag_i225}
\end{figure*}

In this part we fix the following set of parameters:
\begin{itemize}
\item $\Delta z_{\text{u}}= 0.02$.
\item $N_{\text{r}}=200 \ 000$ ($\sim 0.55$ gal.arcmin$^{-2}$).
\item $N_{\text{u}}= 2 \ 000$ ($\sim 0.55$ $10^{-2}$ gal.arcmin$^{-2}$).
\end{itemize}
\noindent We then vary the magnitude of the unknown sample to be: 
\begin{itemize}
\item $i_{\text{u}}<22.5$.
\item $22.5 \leq i_{\text{u}}<23$.
\item $23 \leq i_{\text{u}}<23.5$.
\item $23.5 \leq i_{\text{u}}<24$.
\item $24 \leq i_{\text{u}}<24.5$.
\item $24.5 \leq i_{\text{u}}<25$. 
\end{itemize}
\noindent The unknown samples selected here, correspond to the magnitude bins visible on Figure \ref{fig_vary_with_mag}, their redshift distribution are visible on the bottom panels of this figure.

This choice of binning allows to study the evolution of clustering redshift accuracy with fainter magnitude and when the unknown sample is incomplete.

For each of these 6 samples we build tomographic subsamples according to $\Delta z_{\text{u}}= 0.02$ and $N_{\text{u}}= 2 \ 000$. Therefore, due to this choice of criteria and depending on the magnitude bin there are some regions in redshift where there is no data. 

For each magnitude bin we compute the clustering redshift distribution of all subsamples. The cluster-$z$/spectro-$z$ plot for the bin $i_{\text{u}}<22.5$ is shown in Figure \ref{fig_vary_mag_i225}. The right panel shows the clustering versus spectroscopic redshifts for $i_{\text{u}}<22.5$. The dashed area beyond $\bar{z}_{spec}=1.2$ corresponds to redshifts where there is no galaxy available from the unknown sample. For each clustering redshift distribution (vertical lines) we compute the mean cluster-$z$ $\bar{z}_{clust}$. Then we compare this value to the mean spectroscopic redshift $\bar{z}_{spec}$ from the selected tomographic sample $\Delta \bar{z}= \bar{z}_{clust} - \bar{z}_{spec} $, visible on the top left panel. From this distribution, we compute an estimate of the standard deviation as: $\sigma = \sigma_{\Delta \bar{z}} / (1+\bar{z}_{spec})$, where $\sigma_{\Delta \bar{z}}= 1.48 \times \text{median}(| \Delta \bar{z} |)$. The bottom left panel shows the histogram of redshift densities where no signal is expected. $\mu_{\text{noise}}$ and $\sigma_{\text{noise}}$ then reflect the background noise. We also compute the mean signal to noise ratio per magnitude bin, SNR defined as: $\langle$SNR$\rangle=$ mean(SNR$_i$), where SNR$_i$ is the signal to noise ratio of each clustering redshift distribution estimated by fitting a Gaussian and taking the ratio between the amplitude and its uncertainty.

We give all plots as complementary online supplement for the reader and we summarize all results in Table \ref{table_summary_vary_magi}. From this table it appears that the clustering based redshift inference does not intrinsically depend on the i-band magnitude of the unknown population at a level higher than few 0.1\% in the measurement of the mean redshift.

\begin{table}
\begin{tabular}{llllll}
\hline
\multicolumn{1}{|c|}{Magnitude} & \multicolumn{1}{|c|}{$\langle \Delta \bar{z} \rangle$} &  \multicolumn{1}{|c|}{$\sigma$} & \multicolumn{1}{|c|}{$\mu_{\text{noise}}$}   &  \multicolumn{1}{|c|}{$\sigma_{\text{noise}}$} &  \multicolumn{1}{|c|}{$\langle$SNR$\rangle$}  \\
\hline
\multicolumn{1}{|c|}{$i_{\text{u}}<22.5$} & \multicolumn{1}{|c|}{0.001} &  \multicolumn{1}{|c|}{0.002} & \multicolumn{1}{|c|}{0.001} & \multicolumn{1}{|c|}{0.007} & \multicolumn{1}{|c|}{20}  \\
\hline
\multicolumn{1}{|c|}{$22.5 \leq i_{\text{u}}<23$} & \multicolumn{1}{|c|}{0.001} &  \multicolumn{1}{|c|}{0.002} & \multicolumn{1}{|c|}{0.001} & \multicolumn{1}{|c|}{0.007} & \multicolumn{1}{|c|}{19}  \\
\hline
\multicolumn{1}{|c|}{$23 \leq i_{\text{u}}<23.5$} & \multicolumn{1}{|c|}{0.001} &  \multicolumn{1}{|c|}{0.002} & \multicolumn{1}{|c|}{0.001} & \multicolumn{1}{|c|}{0.007} & \multicolumn{1}{|c|}{19}  \\
\hline
\multicolumn{1}{|c|}{$23.5 \leq i_{\text{u}}<24$} & \multicolumn{1}{|c|}{0.001} &  \multicolumn{1}{|c|}{0.002} & \multicolumn{1}{|c|}{0.001} & \multicolumn{1}{|c|}{0.007} & \multicolumn{1}{|c|}{20}  \\
\hline
\multicolumn{1}{|c|}{$24 \leq i_{\text{u}}<24.5$} & \multicolumn{1}{|c|}{0.001} &  \multicolumn{1}{|c|}{0.002} & \multicolumn{1}{|c|}{0.001} & \multicolumn{1}{|c|}{0.008} & \multicolumn{1}{|c|}{19}  \\
\hline
\multicolumn{1}{|c|}{$24.5 \leq i_{\text{u}}<25$} & \multicolumn{1}{|c|}{0.001} &  \multicolumn{1}{|c|}{0.002} & \multicolumn{1}{|c|}{0.001} & \multicolumn{1}{|c|}{0.007} & \multicolumn{1}{|c|}{22}  \\

\hline
\end{tabular}
\caption{This table summarises the accuracy of the clustering-based redshift reconstruction on the mean redshift when varying the $i$-band magnitude from 22.5 to 25 by step of 0.5. From these tests it appears that the method is stable and does not depend on the $i$-band magnitude at a level higher than a few 0.1\%. }
\label{table_summary_vary_magi}
\end{table}

\subsection{Evolution with the number of objects in both the unknown and reference sample:  }

In the first part we fix the following set of parameters:
\begin{itemize}
\item $\Delta z_{\text{u}}= 0.02$.
\item $N_{\text{r}}=200 \ 000$ ($\sim 0.004$ gal.arcmin$^{-2}$.$(\delta z_{\text{r}}/0.01)^{-1}$).
\item $i_{\text{u}}<22.5$.
\end{itemize}
\noindent We then vary the number of unknown objects in each tomographic bin to be:
\begin{itemize}
\item $N_{\text{u}}= 2 \ 000$ ($\sim 0.55$ $10^{-2}$ gal.arcmin$^{-2}$).
\item $N_{\text{u}}= 1 \ 500$ ($\sim 0.41$ $10^{-2}$ gal.arcmin$^{-2}$).
\item $N_{\text{u}}= 1 \ 000$ ($\sim 0.27$ $10^{-2}$ gal.arcmin$^{-2}$).
\item $N_{\text{u}}=  500$ ($\sim 0.14$ $10^{-2}$ gal.arcmin$^{-2}$).
\item $N_{\text{u}}= 250$ ($\sim 0.07$ $10^{-2}$ gal.arcmin$^{-2}$). 
\end{itemize} 

Then, to study the evolution in the number of reference sources we fixed $N_{\text{u}}= 2 \ 000$} and start from the optimistic reference sample. We go to lower and lower pessimistic spectroscopic densities :
\begin{itemize}
\item $N_{\text{r}}=200 \ 000$ ($\sim 0.004$ gal.arcmin$^{-2}$.$(\delta z_{\text{r}}/0.01)^{-1}$). \item $N_{\text{r}}=150 \ 000$ ($\sim 0.003$ gal.arcmin$^{-2}$.$(\delta z_{\text{r}}/0.01)^{-1}$). 
\item $N_{\text{r}}=100 \ 000$ ($\sim 0.002$ gal.arcmin$^{-2}$.$(\delta z_{\text{r}}/0.01)^{-1}$). \item $N_{\text{r}}=50 \ 000$ ($\sim 0.001$ gal.arcmin$^{-2}$.$(\delta z_{\text{r}}/0.01)^{-1}$).
\item $N_{\text{r}}=25 \ 000$ ($\sim 0.0005$ gal.arcmin$^{-2}$.$(\delta z_{\text{r}}/0.01)^{-1}$).
\end{itemize}

\begin{figure}
\begin{center}
\includegraphics[scale=0.3]{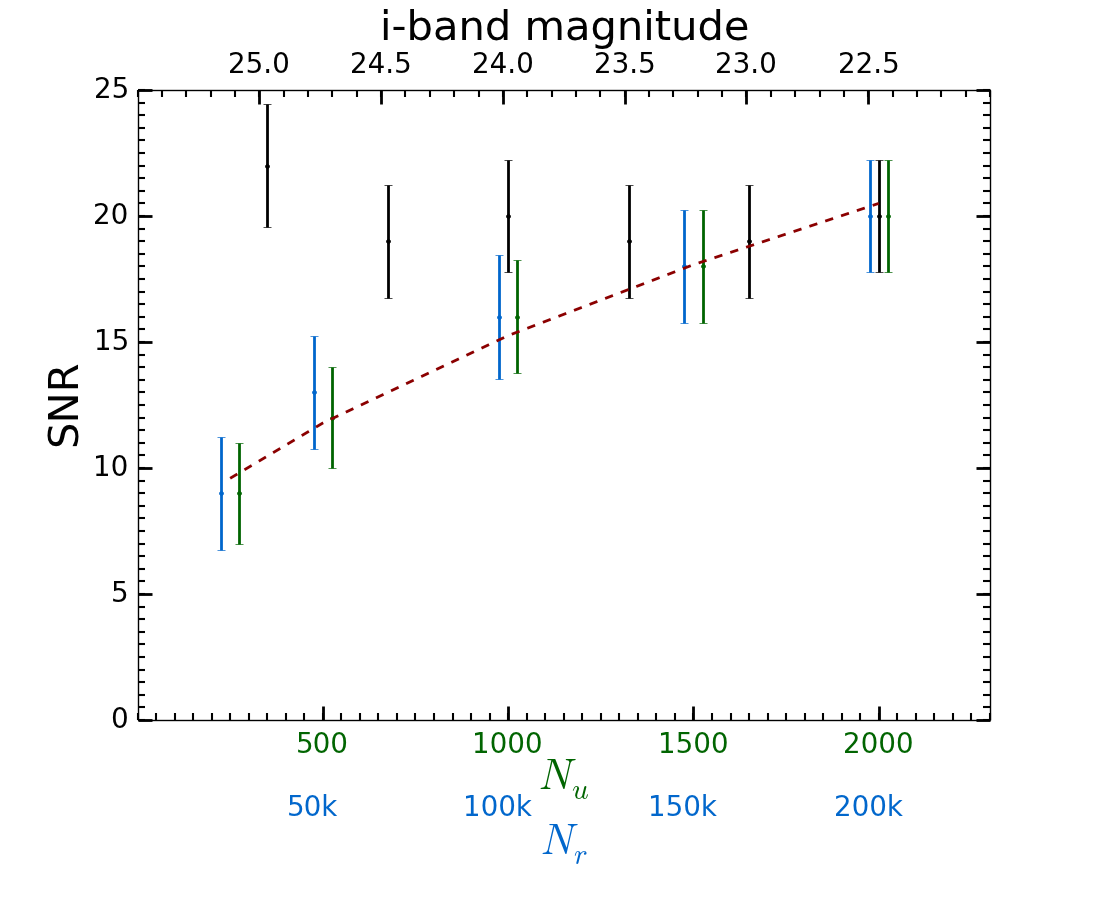}
\end{center}
\caption{Evolution of the signal-to-noise ratio of the clustering-based redshift when varying the $i$-band magnitude of the unknown sample (black) the number of unknown objects (green) and the number of objects in the reference sample (blue). As expected the SNR evolves as the square root (red dashed line) of the number of pairs in the reference and unknown populations.}
\label{fig_SNR}
\end{figure}

\noindent In both cases we follow the same procedure than in Section~\ref{sec:vary_with_mag} and summarise the results in Table \ref{table_summary_vary_Nu}. Figure \ref{fig_SNR} shows that the $\langle$SNR$\rangle$ evolves as $\sqrt{N_{\text{u}} N_{\text{r}}}$. This is in agreement with the expected behaviour of the integrated crosscorrelation function. Indeed a crosscorrelation signal evolves as the number of pairs of objects $N_{\text{u}} N_{\text{r}}$ while in regions where there is no signal, the noise is stochastic and is expected to evolve as $\sqrt{N_{\text{u}} N_{\text{r}}}$.

\begin{table}
\begin{tabular}{llllll}
\hline
\multicolumn{1}{|c|}{Number of objects} & \multicolumn{1}{|c|}{$\langle \Delta \bar{z} \rangle$} &  \multicolumn{1}{|c|}{$\sigma$} & \multicolumn{1}{|c|}{$\mu_{\text{noise}}$}   &  \multicolumn{1}{|c|}{$\sigma_{\text{noise}}$} &  \multicolumn{1}{|c|}{$\langle$SNR$\rangle$}  \\
\hline
\multicolumn{1}{|c|}{$N_{\text{u}}=2000$ } & \multicolumn{1}{|c|}{0.001} &  \multicolumn{1}{|c|}{0.002} & \multicolumn{1}{|c|}{0.001} & \multicolumn{1}{|c|}{0.007} & \multicolumn{1}{|c|}{20} \\
\hline
\multicolumn{1}{|c|}{$N_{\text{u}}=1500$} & \multicolumn{1}{|c|}{0.001} &  \multicolumn{1}{|c|}{0.002} & \multicolumn{1}{|c|}{0.001} & \multicolumn{1}{|c|}{0.008} & \multicolumn{1}{|c|}{18}  \\
\hline
\multicolumn{1}{|c|}{$N_{\text{u}}=1000$} & \multicolumn{1}{|c|}{0.001} &  \multicolumn{1}{|c|}{0.002} & \multicolumn{1}{|c|}{0.001} & \multicolumn{1}{|c|}{0.010} & \multicolumn{1}{|c|}{16}  \\
\hline
\multicolumn{1}{|c|}{$N_{\text{u}}=500$} & \multicolumn{1}{|c|}{-0.017} &  \multicolumn{1}{|c|}{0.002} & \multicolumn{1}{|c|}{0.001} & \multicolumn{1}{|c|}{0.013} & \multicolumn{1}{|c|}{12} \\
\hline
\multicolumn{1}{|c|}{$N_{\text{u}}=250$} & \multicolumn{1}{|c|}{-0.077} &  \multicolumn{1}{|c|}{0.003} & \multicolumn{1}{|c|}{0.001} & \multicolumn{1}{|c|}{0.019} & \multicolumn{1}{|c|}{9} \\
\hline
\hline
\multicolumn{1}{|c|}{$N_{\text{r}}=200k$} & \multicolumn{1}{|c|}{0.001} &  \multicolumn{1}{|c|}{0.002} & \multicolumn{1}{|c|}{0.001} & \multicolumn{1}{|c|}{0.007} & \multicolumn{1}{|c|}{20} \\
\hline
\multicolumn{1}{|c|}{$N_{\text{r}}=150k$} & \multicolumn{1}{|c|}{0.001} &  \multicolumn{1}{|c|}{0.002} & \multicolumn{1}{|c|}{0.001} & \multicolumn{1}{|c|}{0.007} & \multicolumn{1}{|c|}{18}\\
\hline
\multicolumn{1}{|c|}{$N_{\text{r}}=100k$} & \multicolumn{1}{|c|}{0.001} &  \multicolumn{1}{|c|}{0.002} & \multicolumn{1}{|c|}{0.001} & \multicolumn{1}{|c|}{0.009} & \multicolumn{1}{|c|}{17} \\
\hline
\multicolumn{1}{|c|}{$N_{\text{r}}=50k$} & \multicolumn{1}{|c|}{0.002} &  \multicolumn{1}{|c|}{0.002} & \multicolumn{1}{|c|}{0.002} & \multicolumn{1}{|c|}{0.010} & \multicolumn{1}{|c|}{14}\\
\hline
\multicolumn{1}{|c|}{$N_{\text{r}}=25k$} & \multicolumn{1}{|c|}{0.002} &  \multicolumn{1}{|c|}{0.002} & \multicolumn{1}{|c|}{0.002} & \multicolumn{1}{|c|}{0.018} & \multicolumn{1}{|c|}{9}\\

\hline
\end{tabular}
\caption{This table summarises the accuracy of the clustering based redshift reconstruction on the mean redshift when varying the number of unknown objects in each selected bins for $N_{\text{u}}=2000$ down to $250$ and when varying the total number of reference objects: $N_{\text{r}}=200\text{k}$ down to $25\text{k}$.}
\label{table_summary_vary_Nu}
\end{table}

\FloatBarrier
\subsection{Evolution with the width of the unknown distribution: $\Delta z_{\text{u}}$}
\label{sec:vary_with_deltaz_u}

In this part we fix the following set of parameters:
\begin{itemize}
\item $N_{\text{r}}=200 \ 000$ ($\sim 0.004$ gal.arcmin$^{-2}$.$(\delta z_{\text{r}}/0.01)^{-1}$).
\item $N_{\text{u}}= 2 \ 000$ ($\sim 0.55$ $10^{-2}$ gal.arcmin$^{-2}$).
\item $i_{\text{u}}<22.5$.
\end{itemize}
\noindent Finally, we vary the redshift width of the unknown sample to be:  $\Delta z_{\text{u}}= 0.02$, $0.04$, $0.08$, $0.16$, $0.32$, $0.64$.

\begin{figure}
\begin{center}
\includegraphics[scale=0.2]{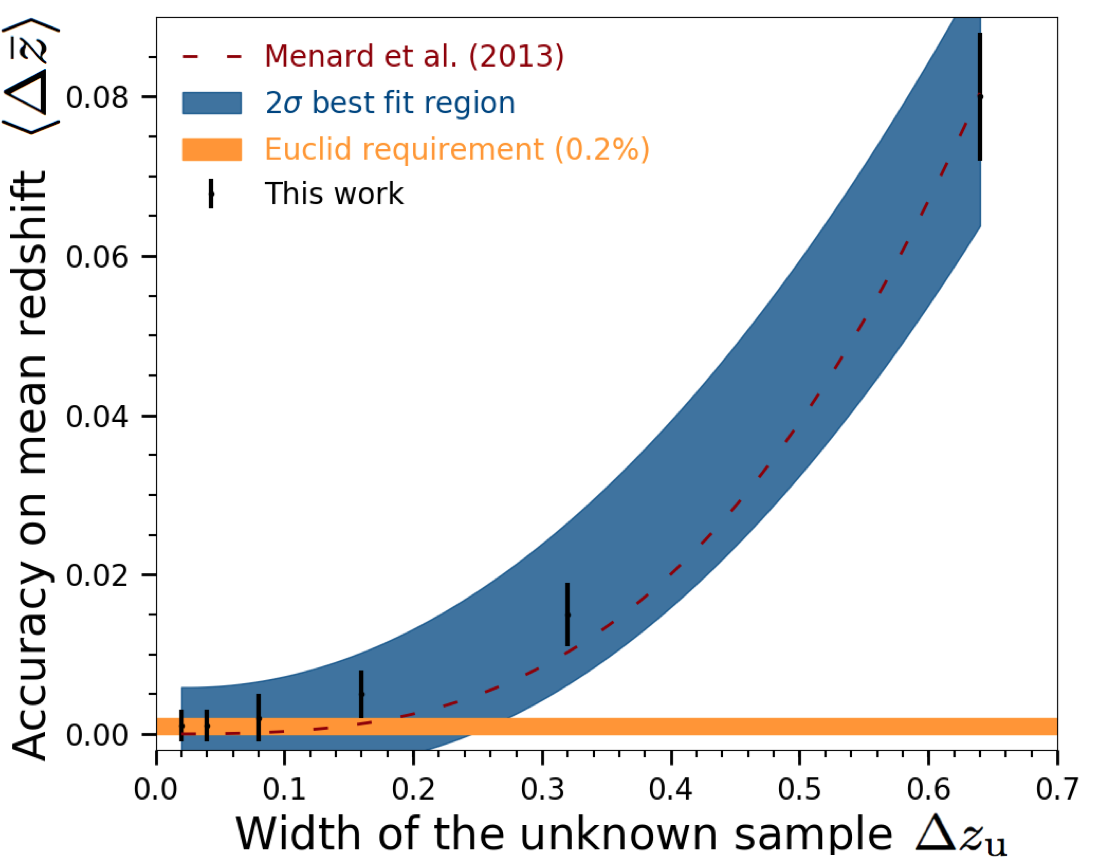}
\end{center}
\caption{Offset in the mean clustering based redshift, when varying the width of the unknown redshift distribution $\Delta z_{\text{u}}$ from $0.02$ to $0.64$. This plot also shows the 2$\sigma$ region around the best fit of the measurement (blue), the Euclid requirement of $0.2\%$ accuracy (yellow) and the expected offset considerring a linear evolution with redshift of the galaxy to dark matter bias (red dashed).}
\label{fig_vary_deltaz_u}
\end{figure}
\noindent Wider distributions progressivly break our hypothesis on the evolution of the unknown clustering amplitude: $\deriv \beta_{\text{u}} / \deriv z = 0$. This can also be seen as a photometric redshift selection with $\Delta z_{\text{u}} \sim 2 \sigma_{\text{phot}}$ where $\sigma_{\text{phot}}$ is the photo-z scatter. This is correct considering no or few catastrophic failures. The results are shown in Figure \ref{fig_vary_deltaz_u} where the red dashed line shows the expected offset considerring a linear evolution with redshift of the galaxy to dark matter bias. This is done assuming a Gaussian redshift distribution following \cite{menard_2013}. This confirm that working with localised unknown samples reduced the non linear evolution of the galaxy to dark matter bias to a linear evolution. The agreement between our measurements and the theoretical expectations also shows that if a survey does not allow to preselect localised distributions narrower than $\Delta z_{\text{u}}=0.2$ then considering no evolution of the galaxy to dark matter bias does not allow to reach $0.2\%$ accuracy in the mean redshifts. Instead, it is highly recommended to consider a linear evolution.

\subsection{Requirements for clustering redshift}

Based on these studies we can now higlight the statistical requirement on the clustering redshift method to reach $\langle \Delta \bar{z} \rangle \sim 0.001 \times (1.+z)$ in few photo-z bins. Then depending on the possibility to preselect narrow distributions one can consider sufficiant to assume no or linear evolution for the galaxy to dark matter bias in the measurement (see Section \ref{sec:vary_with_deltaz_u}). Let's consider a realistic reference sample like the BOSS DR12Q \citep{paris_qso_dr12}. We note that there is at least 2000 QSOs over 9376 deg$^2$ per $\delta z = 0.05$ from redshift 0.5 to 4. This gives an estimate of the available reference density of: 
\begin{equation}
10^{-5} \ \text{qso.arcmin}^{-2}.\left(  \frac{\delta z}{0.01}\right)^{-1} \ .
\end{equation}
In the previous study, we reached the desired accuracy of $0.1\%$ on the bias with $N_{\text{r}}$=100k, $N_{\text{u}}$=2000 over 100 deg$^{2}$. The reference density was then:
\begin{equation}
2.10^{-3} \ \text{gal.arcmin}^{-2}.\left(  \frac{\delta z}{0.01}\right)^{-1} \ ,
\end{equation}
while the unknown one was:
\begin{equation}
5.10^{-3} \ \text{gal.arcmin}^{-2} \ .
\end{equation}
Using BOSS as reference sample implies a requirement on the minimal unknown density of galaxies of $5 \ \text{gal.arcmin}^{-2}$.  Consequently a BOSS-like spectroscopic survey over an area of $9000 \ \text{deg}^{2}$ with $\sim 200$ reference sources per reference slices $\delta z_{\text{r}}=0.005$ allows to reach $\langle \Delta \bar{z} \rangle \sim 0.001 \times (1.+z)$ for tomographic bins.

\section{Individual redshift measurement}
\label{sec:individual_redshift}

\noindent This section aims at demonstrating that it is possible to get accurate PDF(z) measurement for each individual galaxy of the unknown population. Following \cite{scottez_vipers_2016} we present a straightforward approach based on colours.

In the general case the sample distribution is the sum of all distinct PDFs:
\begin{equation}
\frac{\deriv N}{\deriv z}= \sum_i^N \text{PDF}_{\text{i}} (z) \ .
\end{equation}
In the limit case where $\frac{\deriv N}{\deriv z} = \delta^{\text{D}} (z -z_{\text{0}})$, then:
\begin{equation}
\text{PDF}_{\text{i}} (z) = \frac{1}{N} \times  \delta^{\text{D}} (z -z_{\text{0}}) \ .
\end{equation}
When the redshift distribution is not a $\delta^{\text{D}}(z)$ function, this approximation will lead to an error in the measurement. The narrower the distribution, the smaller the error.

In this section we quantify the accuracy of individuals clustering redshifts by computing: the bias$= (  z_{\text{clust}}  -  z_{\text{spec}} ) / (1 + z_{\text{spec}} )$, the outlier fraction $\eta= |  z_{\text{clust}}  -  z_{\text{spec}} |/ (1 + z_{\text{spec}} ) > 0.15  $ and the scatter $\sigma= 1.48 \times  median(|  z_{\text{clust}}  -  z_{\text{spec}} |)/ (1 + z_{\text{spec}} )$.

\subsection{Sampling the true colour space}

\begin{figure}
\begin{center}
\includegraphics[scale=0.35]{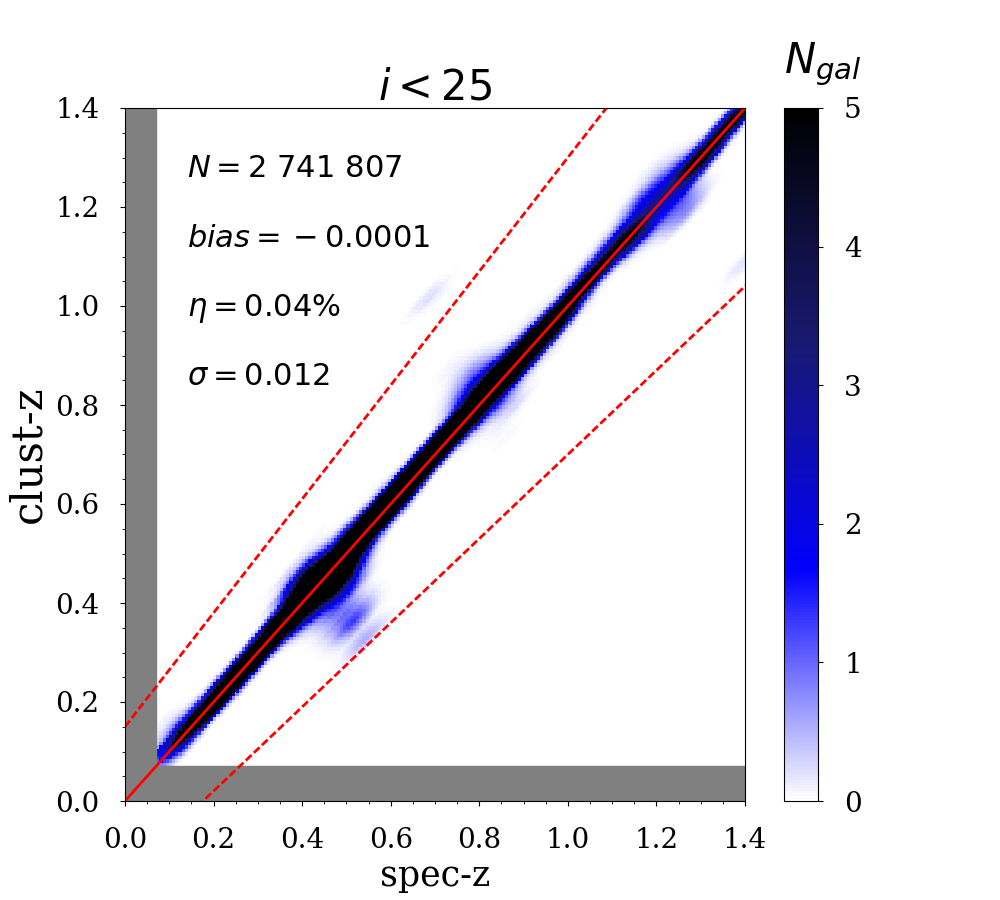}
\end{center}
\caption{Individual clustering redshifts as a function of the spectroscopic redshifts for each galaxy when using perfect photometry. The red dashed lines correspond to the limit beyond which the redshift of a galaxy is considered as catastrophic.}
\label{fig_cluster_spec_z_plane_indiv}
\end{figure}

To preselect narrow redshift distribution, we subsample the parent unknown sample defined in Section \ref{sec_unk_sample} based on galaxy colours using: u, g, r, i, z, Y, J, H photometric bands. One can note that these values are true values since there is no error on magnitude within the MICE2 simulation. Each of these samples correspond to a 7-dimensional colour space volume element of width $\Delta_{\text{colour}}=0.1$. We can then measure the clustering redshift distribution of these samples with $N_{\text{u}} \geq 1000$. Due to this selection the fraction of lost objects is important and is around $67\%$. Nevertheless, this effect is expected to decrease with the increasing size of the unknown sample.\newline
\indent Considering these distributions are sufficiently narrow, we take the median of each distribution and claim that this is our estimate of the individual redshift for all galaxies in this volume element. Applying this on the full color space we can estimate the redshift of each galaxy (see Figure \ref{fig_cluster_spec_z_plane_indiv}). This gaussian kernel density map shows the clustering redshift as a function of the spectroscopic redshift for each galaxy down to $i<25$.\newline
\indent We remind the reader that here we are using perfect photometry. We will explore a realistic case in the next section. However this study demonstrate that our clustering based redshift estimator is not intrinsically biased at a level higher than $0.01\%$.

\subsection{Sampling the real colour space}
\label{sec:individual_redshift_with_mag_err}

\FloatBarrier
\begin{figure}
\begin{center}
\includegraphics[scale=0.35]{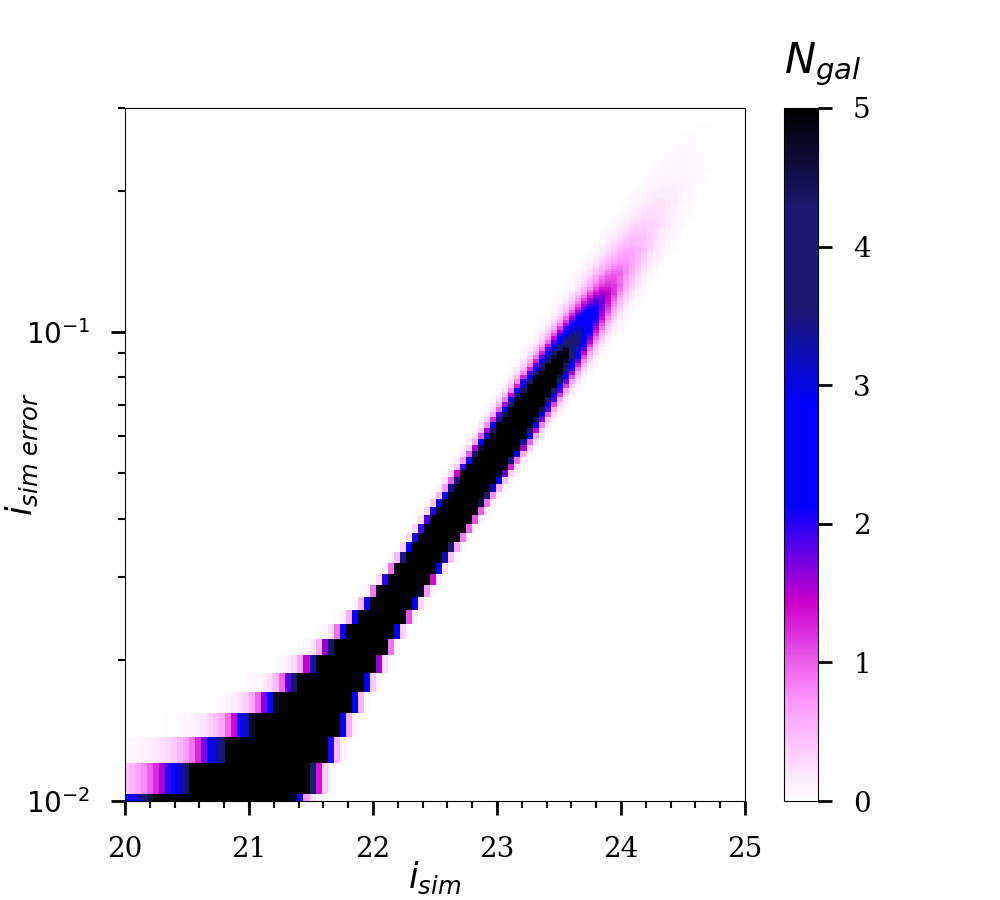}
\end{center}
\caption{Simulated error for the $i$-band versus the true magnitude. This takes into account a gaussian random error on the true flux and a gaussian random error from the sky background through 2 diameter aperture.}
\label{fig_i_sim_err}
\end{figure}

In this section we added realistic error on the magnitude. To do so, we simulate the photometry by assuming megacam instrumental characteristics for $ugriz$. For NIR depth we assum the WIRCam instrument characteristics. The exposure times are tuned to match the following depths at $5\sigma$ for $2''$ diameter apertures $u=24.2$, $g=24.5$, $r=23.9$, $i=23.6$, $z=23.4$, $Y=23$, $J=23$, $H=23$ and modeling the error as a gaussian random error on the true flux and a gaussian random background noise from the sky from 2 diameter aperture. Figure \ref{fig_i_sim_err} shows the variation with magnitude of the resulting error for the $i$-band.

\begin{figure}
\begin{center}
\includegraphics[scale=0.35]{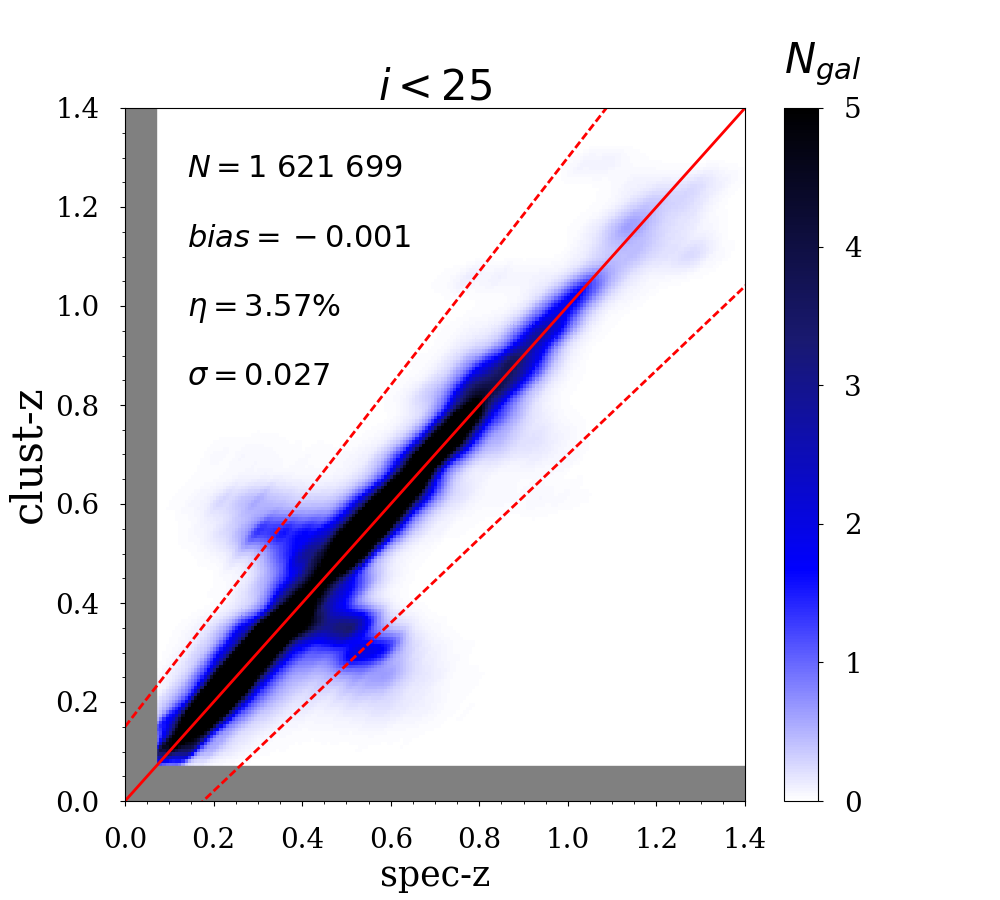}
\end{center}
\caption{Density map of individual clustering redshift as a function of the spectroscopic redshift considering realistic error on magnitude measurements. The red dashed lines correspond to the limit beyond which the redshift of a galaxy is considered as catastrophic.}
\label{fig_cluster_spec_z_plane_indiv_magerr}
\end{figure}

Adding noise to magnitudes induces a scatter of galaxies in colour space. This leads on average to a decrease in the number of galaxies per cell. For this reason we choose to use a resolution of $\Delta_{\text{col}}=0.15$. Then, we apply the same procedure than in Section \ref{sec:individual_redshift}. We show the resulting individual clustering redshifts compare to their true spectroscopic redshifts in Figure~\ref{fig_cluster_spec_z_plane_indiv_magerr}. These objects have $i_{\textit{u}}<25$ with a median at $i_{\text{u}}=23$. While the quality of these measurements obviously degrades with respect to Figure \ref{fig_cluster_spec_z_plane_indiv} they are still quite competitive with photometric redshifts goals of next generation of cosmological surveys in terms of scatter, outlier rate and bias. At this point it seems important to remind the reader that this approach is independent from the photometric redshifts procedure since the colour information is only used in the preselection step and not to extract the redshift information. The redshift information come only from the clustering of objects. The only observables used to extract this information are then the right ascension (RA) and the declination (DEC) of the unknown population.

The advantage of this procedure is threefold: i)  it allows the use of clustering redshifts for any field in extragalactic astronomy, ii) it allows the possibility to combine photometric and clustering based redshifts to get an improved redshift estimation,  iii) it allows the use of cluster-$z$s to define tomographic bins for weak lensing. We explore this last option in the next section.

\begin{figure}
\begin{center}
\includegraphics[scale=0.33]{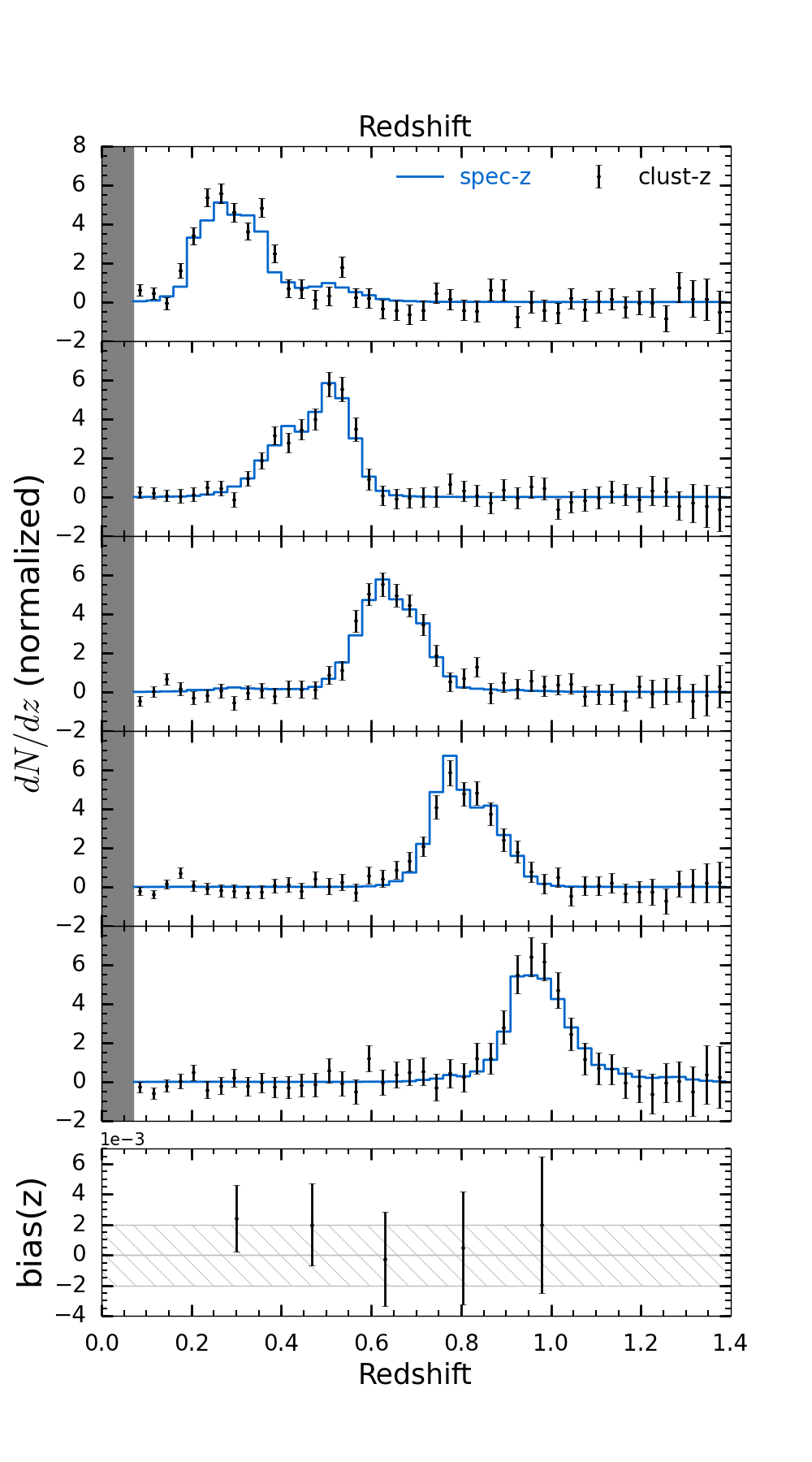}
\end{center}
\caption{True and measured redshift distributions of the five clustering redshift selected tomographic bins. The lower panel shows the bias of each bin (black) while the dashed area corresponds to the Euclid requirements. Please note that here the bias is estimated on the mean redshift of the distribution.}
\label{fig_WL_tomo_bins}
\end{figure}


\subsection{Clustering redshifts based tomographic sampling}
\label{sec:individual_redshift_with_mag_err}

Finally, we investigate the accuracy in the mean of the redshift distribution one can reach in the context of tomographic weak lensing. To reduce the effect of the evolution of the galaxy to dark matter bias of the unknown sample, we selected 30 clustering redshifts bins from 0.2 to 1 based on individual estimates from Figure \ref{fig_cluster_spec_z_plane_indiv_magerr}. For each of these 30 bins we re-measured the redshift distribution to benefit from the increase in $N_{\text{u}}$. These bins are then combined to build 5 tomographic bins as shown in Figure \ref{fig_WL_tomo_bins} where the redshift resolution is $\delta z_{\text{r}}=0.03$ for clarity. This figure shows the true and measured redshift distributions of the five clustering redshift selected tomographic bins. The lower panel shows the bias of each bin (black) while the dashed area corresponds to  $\Delta \langle z \rangle \leq 0.002(1+z)$. Please note that the bias is estimated on the mean redshift of the distribution: $bias= ( \langle z_{\text{clust}} \rangle -  \langle z_{\text{spec}} \rangle ) / (1 + \langle z_{\text{spec}} \rangle )$. We remind the reader that this is done with a sample complete to $i_{\text{u}}=23.5$ due to the simulation. Even if our results in Section~\ref{sec:vary_with_mag} indicate no evolution with the magnitude of the unknown sample, this could be affected by faintest objects at redshifts higher than the simulation limite.

\section{Summary}
\label{sec:summary}

In this study, we aimed at giving a preview of the redshift accuracy one can reach on the redshift estimation using the clustering of galaxies. 

We explored the accuracy of clustering redshifts and we estimate that a density of spectroscopic objects \-- galaxies or QSOs \-- of 10$^{-5}$ sources.arcmin$^{-2}$ per redshift bins of width $\delta_z = 0.01$ allows to reach the $0.1 \%$ accuracy in the estimate of the mean redshift for a galaxy density compatible with next generation cosmological surveys. This number is compatible with the density of Quasi Stellar Objects in BOSS.

We also demonstrated that it is possible to get an estimate of the redshift for each galaxy in a fully independent way from photometric redshifts. The resulting clustering redshifts have a bias$=−0.001$, an outlier fraction of $\eta = 3.57\%$ and a scatter of $\sigma = 0.027$ with a sample complete to $i=23.5$ with maximum magnitude of $i =25$. Clustering redshifts are then competitive compared to photometric redshifts.

These individual measurements: i)  allow the use of clustering redshifts for any field in extragalactic astronomy, ii) allow the possibility to combine photometric and clustering based redshifts to get an improved redshift estimation,  iii) it allows the use of cluster-$z$s to define tomographic bins for weak lensing.

Based on these results we investigated the reachable accuracy in the measurement of the mean of the redshift distribution which is currently still an issue for tomographic weak lensing. In this context, we demonstrated our ability to build 5 clustering redshift selected tomographic bins from redshift 0.2 to 1 with a bias of $0.002$ per bin with a sample complete to $i=23.5$ and $i_{\text{max}} =25$.

Finally, even if Section~\ref{sec:vary_with_mag} indicates no evolution with magnitude and allow us to be confident in the generalisation of this result to fainter complete sample. Nevertheless, this need to be confirmed. Also, the lack of redshift $> 1.4$ which leads to an additional degeneracy in color space resulting in additional noise in the detection. This effect and the use of clustering redshifts for weak lensing will be explored in a future work within the full Flagship simulation \citep{flagship_1}.

\section*{Acknowledgments}

VS acknowledges funding from the Centre National d'Études Spatiales (CNES) through the Convention CNES/CNRS N$^{\text{o}}$ 140988/00 on the scientific development of VIS and NISP instruments and the management of the scientific consortium of the Euclid mission. VS acknowledges the Euclid Consortium and the Euclid Science Working Groups.

The MICE simulations have been developed at the MareNostrum supercomputer (BSC-CNS) thanks to grants AECT-2006-2-0011 through AECT-2015-1-0013. Data products have been stored at the Port d'Informaci\'o Cient\'ifica (PIC), and distributed through the CosmoHub webportal (cosmohub.pic.es). Funding for this project was partially provided by the Spanish Ministerio de Ciencia e Innovacion (MICINN), projects 200850I176, AYA2009- 13936, AYA2012-39620, AYA2013-44327, ESP2013-48274, ESP2014-58384, Consolider-Ingenio CSD2007- 00060, research project 2009-SGR-1398 from Generalitat de Catalunya, and the Ramon y Cajal MICINN program.

\bibliographystyle{mn2e}
\bibliography{biblio}

\label{lastpage}

\end{document}